\documentclass[twocolumn,english,prl,floatfix,citeautoscript,nofootinbib]{revtex4}
\usepackage{amsbsy}
\usepackage{latexsym,epsfig,graphicx}
\usepackage{dcolumn}
\usepackage{subfigure}
\usepackage{comment}
\usepackage{color}
\usepackage[colorlinks,urlcolor=blue,citecolor=blue]{hyperref}
\usepackage{amstext}
\usepackage{amssymb}
\usepackage{setspace}
\usepackage[cp1251]{inputenc}

\begin{document}

\title{Robust Weyl points in a 1D superlattice with transverse spin-orbit
coupling}
\author{Xi-Wang Luo}
\author{Chuanwei Zhang}
\thanks{Corresponding author. \\
Email: \href{mailto:chuanwei.zhang@utdallas.edu}{chuanwei.zhang@utdallas.edu}%
}
\affiliation{Department of Physics, The University of Texas at Dallas, Richardson, Texas
75080-3021, USA}

\begin{abstract}
Weyl points, synthetic magnetic monopoles in the 3D momentum space, are the
key features of topological Weyl semimetals. The observation of Weyl points
in ultracold atomic gases usually relies on the realization of
high-dimensional spin-orbit coupling (SOC) for two pseudospin states (%
\textit{i.e.,} spin-1/2), which requires complex laser configurations and
precise control of laser parameters, thus has not been realized in
experiment. Here we propose that robust Wely points can be realized using 1D
triple-well superlattices (spin-1/three-band systems) with 2D transverse SOC
achieved by Raman-assisted tunnelings. The presence of the third band is
responsible to the robustness of the Weyl points against system parameters
(e.g., Raman laser polarization, phase, incident angle, etc.). Different
from a spin-1/2 system, the non-trivial topology of Weyl points in such
spin-1 system is characterized by both spin vector and tensor textures,
which can be probed using momentum-resolved Rabi spectroscopy. Our proposal
provides a simple yet powerful platform for exploring Weyl physics and
related high-dimensional topological phenomena using high pseudospin
ultracold atoms.
\end{abstract}

\maketitle

\section{Introduction}
Weyl semimetal, an exotic topological
phase of matter, possesses novel quasi-particle excitations behaving as Weyl
fermions in the bulk and intriguing Fermi arcs on the surface~\cite%
{Weyl1929elektron, PhysRevB.83.205101, turner2013beyond, hosur2013recent,
PhysRevLett.114.206602}. The key feature of Weyl semimetal is the appearance
of Weyl points (gapless points in the band structure with linear dispersion
in 3D momentum space) characterized by nontrivial topological invariants~%
\cite{turner2013beyond, hosur2013recent}. Weyl point does not depend on
symmetry except the translational symmetry of the crystal lattice, and is
the most robust degeneracy which can only be gapped out when annihilates
with another Weyl point with opposite topological charge.
Weyl fermions may exhibit
non-trivial electromagnetic responses to external gauge field~\cite{PhysRevB.86.115133,PhysRevB.87.161107,PhysRevB.99.155142}.
Due to the
fundamental importance of Weyl fermions and the potential application of
surface states, significant theoretical and experimental progresses have
been made for exploring Weyl physics in both solid-state materials~\cite%
{PhysRevLett.107.127205, huang2015Weyl, xu2015discovery, lv2015experimental,
parameswaran2014probing, soluyanov2015type, PhysRevLett.117.056805,
chang2016prediction, deng2016experimental, huang2016spectroscopic} and
synthetic systems such as ultracold atomic gases~\cite{PhysRevB.84.165115,
PhysRevLett.108.235301, PhysRevB.91.125438, PhysRevA.85.033640,
PhysRevLett.114.225301, PhysRevA.94.053619, PhysRevA.94.013606,
wang2018dirac}, photonic~\cite{lu2013Weyl, chen2016photonic,
lu2015experimental, PhysRevLett.117.057401,
lin2016photonic,PhysRevA.96.013857} and acoustic crystals~\cite%
{xiao2015synthetic}. In contrast to solid-state materials whose complicated
band structures make the probing of Weyl-fermion topology elusive, synthetic
systems are simple, clean and highly controllable. In particular, recent
experimental realization of 1D and 2D spin-orbit coupling and synthetic gauge field
in ultracold atoms
makes the atomic system one of the most promising platforms for studying
topological effects and novel state of matter~\cite{lin2011spin,
zhang2012collective, qu2013observation, olson2014tunable,
ji2014experimental, wang2012spin, cheuk2012spin, wu2016realization,
meng2016experimental, huang2016experimental, campbell2015itinerant,
luo2016tunable,dalibard2011colloquium,goldman2014light}.

So far most ultracold atom based schemes~\cite{PhysRevB.84.165115,
PhysRevLett.108.235301, PhysRevB.91.125438, PhysRevA.85.033640,
PhysRevLett.114.225301, PhysRevA.94.053619, PhysRevA.94.013606,
wang2018dirac} for realizing Weyl physics rely on the generation of 3D
spin-orbit coupling for two pseudospin states (\textit{i.e.}, spin-1/2) in
either optical lattices or free space, which require complex laser setups.
Furthermore, such Weyl points are usually very sensitive to laser parameters
(e.g., phases, polarizations and incident angles), making the experimental
realization very challenging with current technique. Weyl points were also
proposed in quasi-particle spectra of BCS superfluids with spin-orbit
coupling~\cite{PhysRevLett.115.265304, PhysRevLett.107.195303,
PhysRevLett.112.136402}, but the experimental realization of such superfluid
is difficult due to heating. Finally, probing non-trivial topology of Weyl
points for spin-1/2 systems is another challenging task,
which requires measurements in various spin bases~\cite{PhysRevA.100.063630}
where many precisely controlled pulses are needed.

In this paper, we propose a much simpler scheme to realize robust Weyl
points and probe their non-trivial topology using a 1D superlattice. Instead
of a spin-1/2 system, we consider a three-band (\textit{i.e.}, spin-1) model
using a triple-well superlattice, with neighbor site tunnelings assisted by
three Raman lasers. The Raman-assisted tunnelings also induce momentum
transfer on the transverse plane, leading to 2D SOC in the transverse free
space. Our main results are:

\textit{i}) The three-band system supports two Weyl points, corresponding to
the degeneracy between two-lower and two-upper bands, respectively.
Therefore they cannot annihilate with each other and any change in system
parameters only shifts their positions, leading to the robustness against
variations of laser parameters (e.g., incident angle, intensity, phase,
detuning and polarization). Such robustness originates from the higher
dimensional Hilbert space enabled by the spin-1 system, which reduces the
requirement for precisely controlled SOC for spin-1/2 systems.

\textit{ii}) For any two neighbor bands, the corresponding surface states
would connect the Weyl point to infinite momentum, indicating that there is
a virtual Weyl point (with opposite charge) at infinity. This can also be
seen by the trajectory of the Weyl point, which may annihilate with its
virtual partner only when it is shifted to infinity for certain critical
system parameters. Away from these critical values, the Weyl points persist.

\textit{iii}) Though the Berry flux around the Weyl point possesses monopole
behavior, the spin textures for such spin-1 system is very different from
the spin-1/2 system~\cite{PhysRevLett.114.225301, PhysRevA.94.053619,
PhysRevA.94.013606, wang2018dirac}. Since the spin-1 vector may go into the
Bloch sphere representing the phase space and even vanish by crossing the center, the non-trivial
topology of the Weyl point is characterized not only by the spin vectors,
but also the spin tensors. We also find that
there is a one-to-one correspondence between the non-trivial Chern number
and the number of vanishing points in the spin vector textures around the Weyl point.

\textit{iv}) We propose a simple scheme to detect the non-trivial topology
of Weyl points based on momentum-resolved Rabi spectroscopy and
time-of-flight imaging. Surprisingly, the additional trivial band near the
Weyl points can serve as a reference which greatly simplifies the detection
pulse sequence.

\begin{figure}[t]
\includegraphics[width=1.0\linewidth]{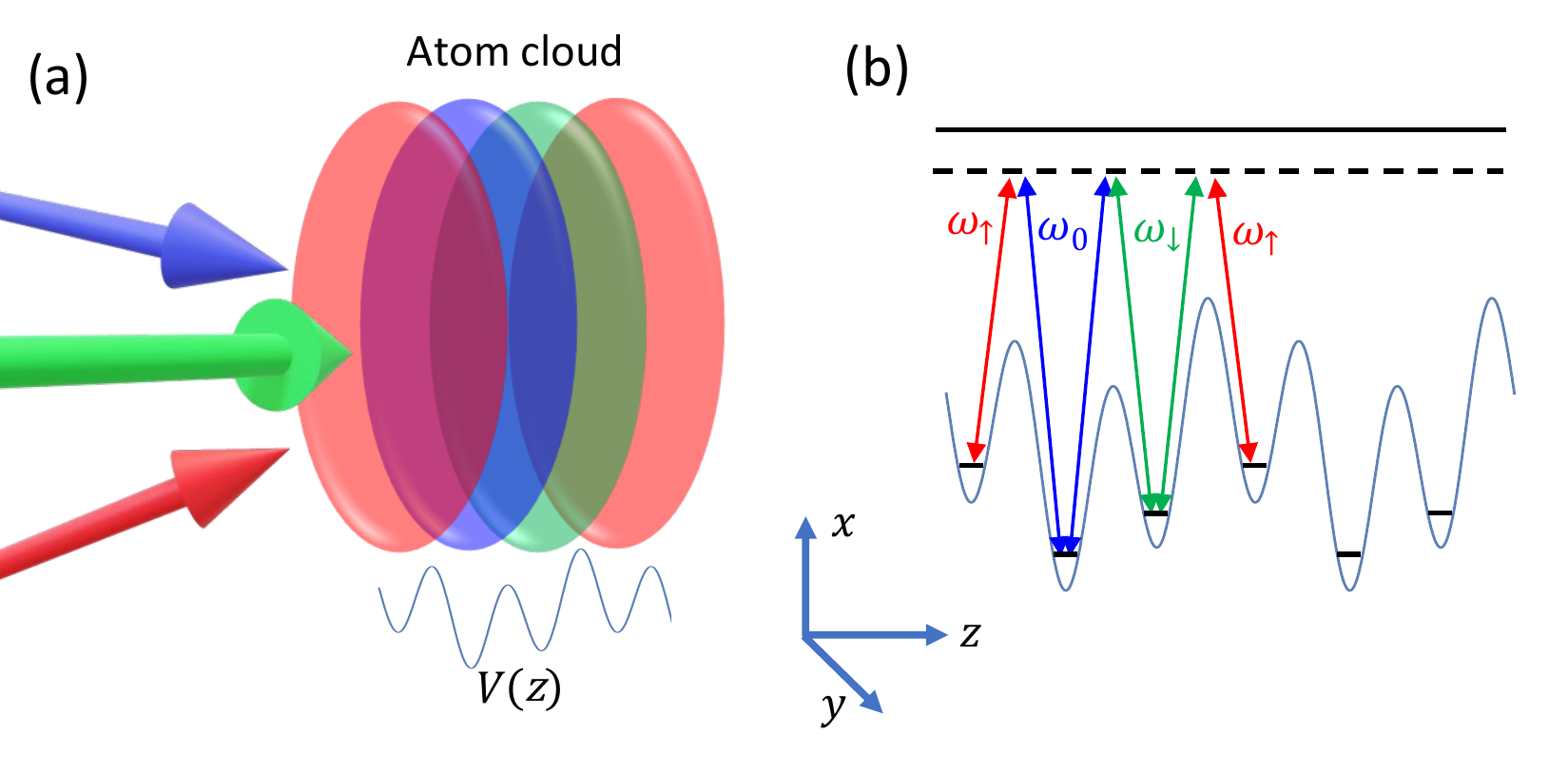}
\caption{(a) Experimental scheme for generating robust Weyl points using a
triple-well superlattice. Three Raman lasers are used to induce the neighbor
site tunnelings. (b) Corresponding level structure and two-photon Raman
transitions in the superlattice.}
\label{fig:sys}
\end{figure}

\section{The model}
We consider a simple experimental setup
shown in Fig.~\ref{fig:sys}a, which contains a 1D superlattice along the $%
z $ direction with 3 sites in each unit cell and is free in 2D $xy$-plane
(i.e., no transverse lattices). The detunings between different sites in the
unit cell are large and the bare tunnelings are suppressed significantly. We
introduce the Raman-assisted tunnelings using three Raman lasers~\cite%
{dalibard2011colloquium, goldman2014light}, with each site in the unit
cell addressed by one and only one Raman laser whose frequency difference is
chosen to match the lattice site detuning [see Fig.~\ref{fig:sys}b]. The
wave vectors should have nonzero components along $z$ to induce momentum kicks
which is needed to generate the
tunneling between neighbour sites. The Raman-assisted transitions acquire transverse momentum kicks
that are determined by the transverse components of the wave vectors.

We adopt the
tight-binding approach and expand the wave function as $|\Psi (\mathbf{r}%
)\rangle =\sum_{j}a_{j}(x,y)|W_{j}(z)\rangle $, with $|W_{j}(z)\rangle $ the
Wannier function for site $j$ in the $z$ direction. The pseudo-spin
operators in each unit cell are denoted as $b_{l,\downarrow }(x,y)=a_{3l}({%
x,y})$, $b_{l,0}(x,y)=a_{3l-1}({x,y})$, $b_{l,\uparrow }(x,y)=a_{3l-2}(x,y)$
with $l$ the unit-cell index. The detunings between them are $\Delta
_{\uparrow }$, $\Delta _{\downarrow }$, and $\Delta _{\uparrow }\pm \Delta
_{\downarrow }$ (see Fig.~\ref{fig:sys}), which are much larger than
the bare nearest neighbor tunneling. Resonance tunnelings between neighbor
lattice sites are induced by three Raman lasers with frequencies $\omega
_{\uparrow }$, $\omega _{0}$, $\omega _{\downarrow }$, satisfying $\omega
_{\uparrow }-\omega _{0}\simeq \Delta _{\uparrow }$, and $\omega
_{\downarrow }-\omega _{0}\simeq \Delta _{\downarrow }$. The pseudo-spin
state $|s\rangle $ is addressed by the laser with frequency $\omega _{s}$ (%
$s=\uparrow ,0,\downarrow $),
which induce both intra- and inter-unit-cell tunnelings.
The Raman-laser wave vectors $\mathbf{K}_s=\mathbf{K}_{z,s}+\mathbf{K}_{\perp,s}$
(with $\mathbf{K}_{z,s}\neq \mathbf{K}_{z,s'}$ for $s\neq s'$)
have nonzero components along both longitudinal
and transverse directions, and the latter induces transverse spin-orbit couplings. The
single-particle Hamiltonian in the rotating frame is
\begin{equation}
\mathcal{H}=\sum_{l,s}\left(\frac{\mathbf{k}_{\bot }^{2}}{2m}+\delta _{s}\right)|l,s\rangle
\langle l,s|+\sum_{l,s;l^{\prime },s^{\prime }}J_{l,s;l^{\prime },s^{\prime
}}|l,s\rangle \langle l^{\prime },s^{\prime }|,
\end{equation}%
where  $|l,s\rangle$ is the single-particle state at unit-cell $l$ with spin $s$,
the non-zero Raman-assisted tunnelings
are $J_{l,0;l,\uparrow }=J_{1}e^{i(\mathbf{K}_{\uparrow}\cdot \mathbf{r}_{l\uparrow}-\mathbf{K}_{0}\cdot \mathbf{r}_{l0})}$, $%
J_{l,\downarrow ;l,0}=J_{2}e^{i(\mathbf{K}_{0}\cdot \mathbf{r}_{l0}-\mathbf{K}_{\downarrow }\cdot \mathbf{r}_{l\downarrow})}$
and $J_{l+1,\uparrow ;l,\downarrow
}=J_{3}e^{i(\mathbf{K}_{\downarrow }\cdot \mathbf{r}_{l\downarrow}-\mathbf{K}_{\uparrow
}\cdot \mathbf{r}_{l\uparrow})}$ with $\mathbf{r}_{ls}=(x,y,z_{ls})$ the coordinates of atoms at
site $(l,s)$.
$\delta _{s}$ is the corresponding detunings and $\mathbf{k}_{\bot }=(k_{x},k_{y})$ is the transverse momentum.

In the quasi-momentum frame after the transformation $|l,s\rangle\rightarrow e^{i\mathbf{K}_{s}\cdot \mathbf{r}_{ls}}|l,s\rangle$,
we obtain the Hamiltonian
\begin{eqnarray}
\mathcal{H} &=&\sum_{l,s}\left[\frac{(\mathbf{k}_{\bot }-\mathbf{K}_{\bot
,s})^{2}}{2m}+\delta _{s}\right]|l,s\rangle \langle l,s|  \nonumber \\
&&+\sum_{l}\big(J_{1}|l,0\rangle \langle l,\uparrow |+J_{2}|l,\downarrow \rangle
\langle l,0| \nonumber \\
&&+J_{3}|l+1,\uparrow \rangle \langle l,\downarrow |+h.c.\big),
\label{Ham_s2}
\end{eqnarray}%
where $%
\mathbf{K}_{\bot ,s}$ corresponds to the transverse momentum kick by the $s$%
-th Raman laser, which gives the 2D spin-orbit coupling strengths in the transverse
direction. The
Hamiltonian in the Bloch momentum space is
\begin{equation}
H_{\mathbf{k}}=\sum_{s}\left[\frac{(\mathbf{k}_{\bot }-\mathbf{K}_{\bot
,s})^{2}}{2m}+\delta
_{s}\right]|s\rangle \langle s|+\sum_{s\neq s^{\prime }}J_{s,s^{\prime }}(\mathbf{k%
})|s\rangle \langle s^{\prime }|.  \label{Ham_1}
\end{equation}%
The
intra-unit-cell couplings are $J_{\uparrow ,0}(\mathbf{k})\equiv J_{0,\uparrow
}^{\ast }(\mathbf{k})=J_{1}$ and $J_{0,\downarrow }(\mathbf{k})\equiv
J_{\downarrow ,0}^{\ast }(\mathbf{k})=J_{2}$, the inter-unit-cell coupling
is $J_{\downarrow ,\uparrow }(\mathbf{k})=J_{\uparrow ,\downarrow }^{\ast }(%
\mathbf{k})=J_{3}e^{ik_{z}}$.
In general, the tunnelling coefficients $%
J_{1,2,3}$ are complex whose phase are determined by the global phases of the Raman
lasers. However, we notice that these phases are unimportant and can be gauged out by absorbing
them in to the definition of the pseudospin state on each site,
and there is no need for fine-tuning of
the Raman-laser phases. We can simply
set $J_{1,2,3}$ to be real (we will assume all $J_{1,2,3}$ positive unless
otherwise state).

\begin{figure}[t]
\includegraphics[width=1.0\linewidth]{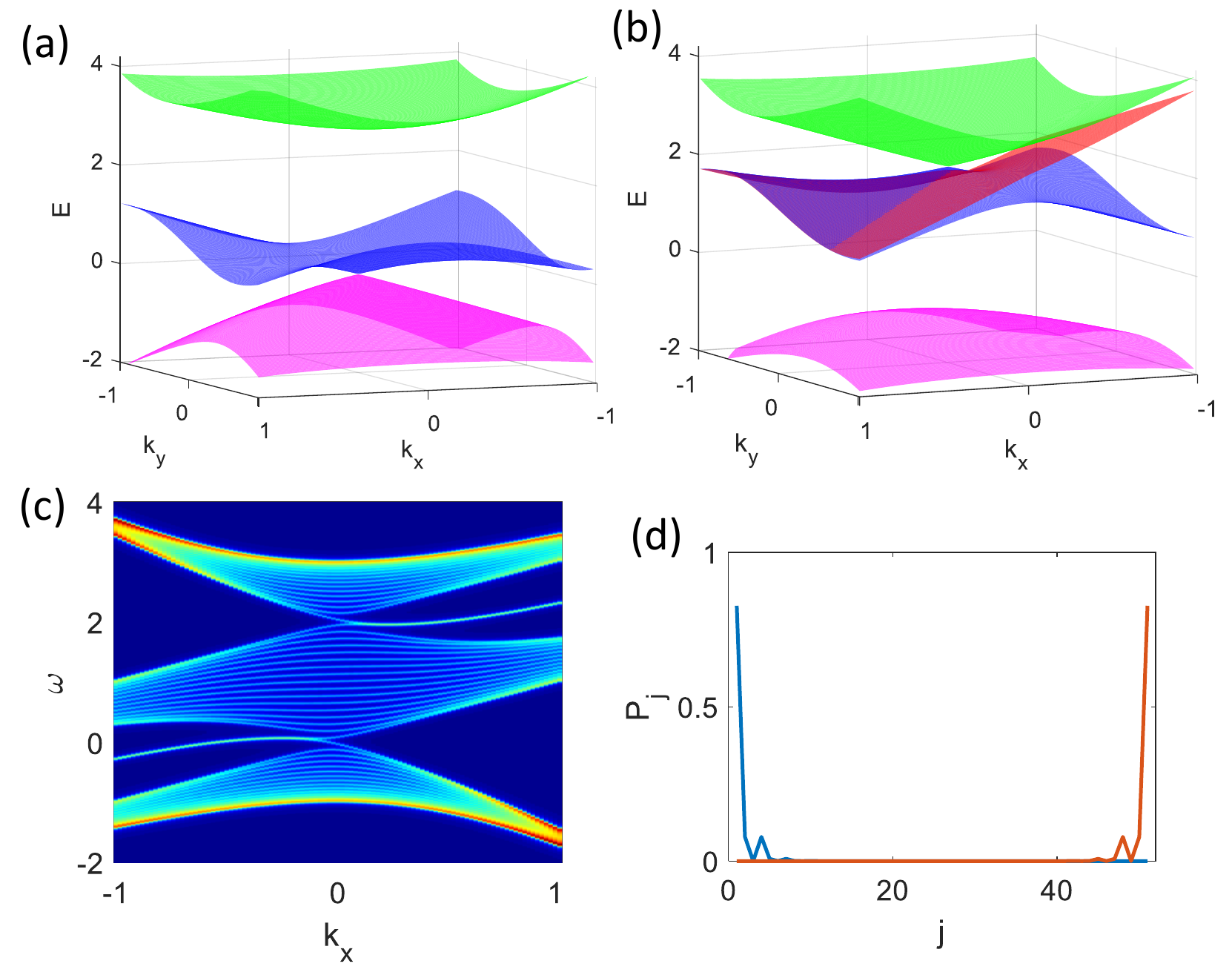}
\caption{(a) Band structure at $k_{z}=0$ with Weyl point $W_{1}$ between two
lower bands. (b) Band structure at $k_{z}=\protect\pi $ with Weyl point $%
W_{2}$ between two upper bands. The mid gap band shows the surface state
dispersion on $z=0$ when open boundary conditions are considered. (c)
Surface spectrum densities $\text{Im}\frac{\protect\gamma }{\protect\pi }%
\frac{1}{\protect\omega -H-i\protect\gamma }$ at different frequencies and
momenta, with $\protect\gamma =0.02$ and $k_{y}=0$. The two surface arcs on
the two surfaces coincide. (d) Typical distribution ($P_{j}$) of the surface
states along the open direction \textit{z}, with $j$ the site index.
$\frac{\mathbf{k}_{\bot}^{2}}{2m}$  is dropped when plotting the band structures. Other
parameters are $J_{1}=J_{2}=J_{3}=1$, $\mathbf{K}_{\bot ,\uparrow }=(-1/2,%
\protect\sqrt{3}/2)$, $\mathbf{K}_{\bot ,0}=(1,0)$, $\mathbf{K}_{\bot
,\downarrow }=(-1/2,-\protect\sqrt{3}/2)$, $\protect\delta _{\uparrow }=%
\protect\delta _{0}=\protect\delta _{\downarrow }=0$.}
\label{fig:band}
\end{figure}

The Hamiltonian in Eq.~(\ref{Ham_1}) supports two robust Weyl points
when all three couplings $J_{1,2,3}$ are
nonzero and the three points $\mathbf{K}_{\bot ,s}=(K_{x,s},K_{y,s})$ are
non-collinear (see Appendix A).
In this paper, we are interested in the Weyl physics
where the three points
$\mathbf{K}_{\bot ,s}=(K_{x,s},K_{y,s})$ form a triangle (i.e., they are
non-collinear). So, we can denote $\mathbf{Q}$ as the triangle's circumcenter and set $\mathbf{Q}$
as the origin of quasi-momentum frame by
a transformation $|l,s\rangle\rightarrow e^{-i\mathbf{Q}\cdot \mathbf{r}_{ls}}|l,s\rangle$,
the Hamiltonian reads
\begin{equation}
H_{\mathbf{k}}=\sum_{s}\left[\frac{(\mathbf{k}_{\bot }-\mathbf{K}'_{\bot ,s})^{2}}{2m}+\delta
_{s}\right]|s\rangle \langle s|
+\sum_{s\neq s^{\prime }}J_{s,s^{\prime }}(\mathbf{k%
})|s\rangle \langle s^{\prime }|
\label{eq:Ham_f}
\end{equation}%
with $\mathbf{K'}_{\bot ,s}=\mathbf{K}_{\bot ,s}-\mathbf{Q}$.
Therefore, we have $|\mathbf{K'}_{\bot ,s}|\equiv k_{R,\bot}$ for
the three spin states $s$.
We will set the momentum and energy unit as
$k_{R,\bot}$ and $\frac{k_{R,\bot}^{2}}{2m}=1$, and consider that the
three points $\mathbf{K'}_{\bot ,s}=(K'_{x,s},K'_{y,s})$ changes on a unit circle
(i.e., the three points $\mathbf{K}_{\bot ,s}=(K_{x,s},K_{y,s})$ changes on a unit circle
centered at $\mathbf{Q}$).
In the following, we will focus on the Hamiltonian Eq.~(\ref{eq:Ham_f}) and drop
the prime symbol in $\mathbf{K'}_{\bot ,s}$ for simplicity.

We denote the two robust Weyl points as $W_{1}$
and $W_{2}$ at $[k_{x}^{W_{1}},k_{y}^{W_{1}},0]$ and $%
[k_{x}^{W_{2}},k_{y}^{W_{2}},\pi ]$. The Weyl point $W_{1}$ corresponds to the degeneracy between
two lower bands while $W_{2}$ for two upper bands. $W_{1}$ and $W_{2}$ are
related to different bands, therefore they cannot annihilate with each
other. Any change in system parameters only shifts the positions of the Weyl
points. The typical band structures are shown in Figs.~\ref{fig:band}a and~%
\ref{fig:band}b, where two Weyl points are clearly shown.
$\frac{\mathbf{k}_{\bot}^{2}}{2m}$ is dropped when plotting the band structures
in Fig.~\ref{fig:band},
our system is not a semimetal and the dispersion relation at higher values of $\mathbf{k}_{\bot}$ goes up in energy.


\begin{figure}[t]
\includegraphics[width=1.0\linewidth]{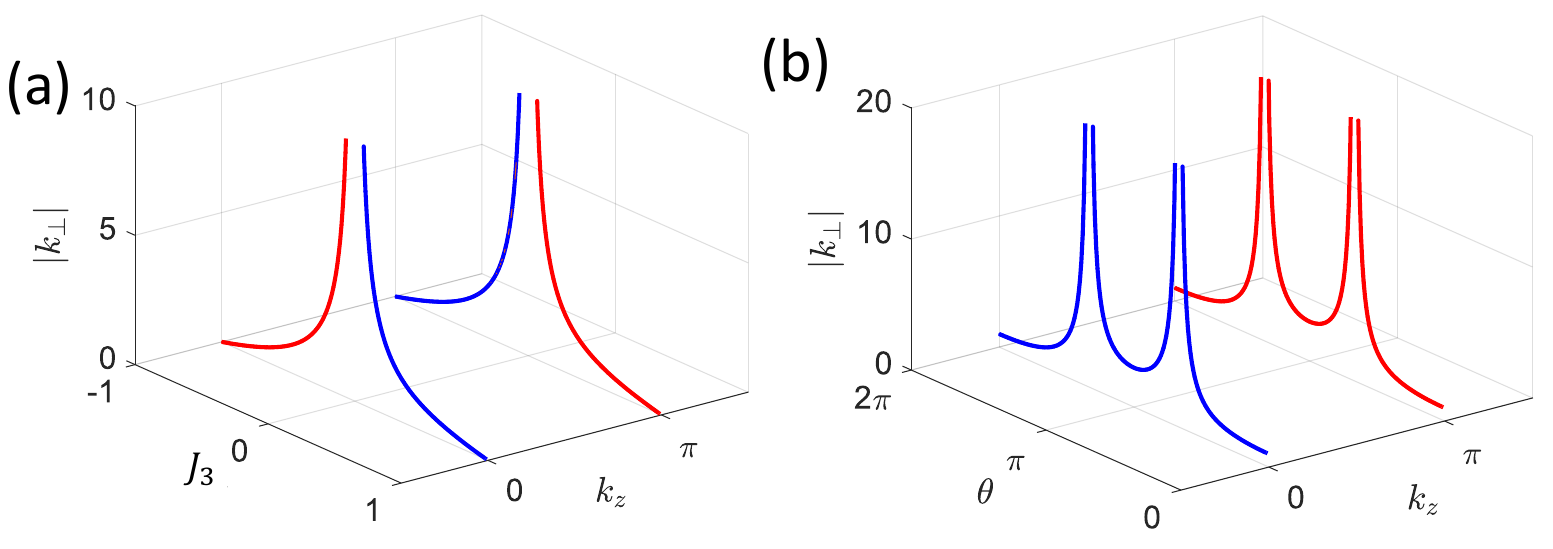}
\caption{(a) The trajectory of Weyl points $W_{1}$ (blue line) and $W_{2}$
(red line) as the inter-unit-cell tunneling varies across zero. $%
J_{1}=J_{2}=1$, $\mathbf{K}_{\bot ,\uparrow }=(-1/2,\protect\sqrt{3}/2)$, $%
\mathbf{K}_{\bot ,0}=(1,0)$, $\mathbf{K}_{\bot ,\downarrow }=(-1/2,-\protect%
\sqrt{3}/2)$. Two Weyl points shift to infinite momenta at $J_{3}=0$. (b)
The trajectories of Weyl points $W_{1}$ (blue line) and $W_{2}$ (red line)
when $\mathbf{K}_{\bot ,0}=(\cos \protect\theta ,\sin \protect\theta )$
rotates in the $k_{x}$-$k_{y}$ plane, $\mathbf{K}_{\bot ,\uparrow }=(-1/2,%
\protect\sqrt{3}/2)$, $\mathbf{K}_{\bot ,\downarrow }=(-1/2,-\protect\sqrt{3}%
/2)$, $J_{1}=J_{2}=2J_{3}=1$. The two Weyl points shift to infinite momenta
at $\protect\theta =2\protect\pi /3$ and $4\protect\pi /3$ where $\mathbf{K}%
_{\bot ,0}$ coincides with $\mathbf{K}_{\bot ,\uparrow }$ and $\mathbf{K}%
_{\bot ,\downarrow }$ respectively.}
\label{fig:traj}
\end{figure}

\section{Surface arcs and Weyl point trajectories}
In
general, Weyl points between any two bands should appear in pairs for a 3D
lattice system because the Brillouin zone is a closed manifold without
boundary~\cite{PhysRevLett.114.225301, PhysRevA.94.053619,
PhysRevA.94.013606, wang2018dirac}. Our system is free in the $xy$ plane,
therefore we could have only one Weyl point between two bands since the
momentum space is an open manifold that may have non-vanishing flux on the
boundary (at infinite $k_{x}$ and $k_{y}$). This can be seen by looking at
the surface arcs which can only start (end) at the Weyl points. In Figs.~\ref%
{fig:band}b and \ref{fig:band}c, we plot the surface arcs with an open
boundary condition along the $z$-direction. Each boundary (left and right)
gives a surface arc which connects the Weyl point to infinite momenta. Shown
in Fig.~\ref{fig:band}d are the distributions of the surface states that
are well localized at the boundary.

There is only one Weyl point between two neighbor bands, which can
annihilate with its virtual partner only when it is shifted to infinity at
certain critical system parameters. The only symmetry required here is the
lattice symmetry along the $z$-direction, thus the Weyl points are very
robust against system disorders. Shown in Fig.~\ref{fig:traj} are the
trajectories of two Weyl points $W_{1}$ and $W_{2}$ as functions of $J_{1,2,3}$
and $\mathbf{K}_{\bot ,s}$. As one of the couplings $J_{1,2,3}$ changes across
the critical value 0 (from positive to negative), two Weyl points first
disappear then reappear at infinity ($\mathbf{k}_{\bot }\rightarrow \infty $%
), with $k_{z}$ changing from 0 ($\pi $) to $\pi $ (0) for Weyl point $W_{1}$
($W_{2}$). The two Weyl points move similarly as $\mathbf{K}_{\bot ,s}-%
\mathbf{K}_{\bot ,s^{\prime }}$ changes across the critical value 0, except
that $k_{z}$ is fixed for both of them. The Raman-laser phases are
irrelevant since the phases of $J_{1,2,3}$ do not affect the band structure
(e.g., the phase of $J_{3}$ only induces a global shift of all bands along $%
k_{z}$). Finally, the pseudospin is represented by different superlattice
sites on the same atomic hyperfine state, making the tunnelings $J_{1,2,3}$
insensitive to laser polarizations.


\section{Berry flux and spin textures}
The topological
properties of the Weyl point can be characterized by the first Chern number~%
\cite{turner2013beyond, hosur2013recent}
\begin{equation}
\mathcal{C}_{n}=\frac{1}{2\pi }\oint_{\mathbf{S}}\mathbf{\nabla }_{\mathbf{k}%
}\times \mathbf{\mathcal{A}}_{n}(\mathbf{k})\cdot d\mathbf{S},
\end{equation}%
where $\mathbf{S}$ is a momentum-space surface enclosing the Weyl point, and
$\mathbf{\mathcal{A}}_{n}(\mathbf{k})=i\langle u_{n}(\mathbf{k})|\mathbf{%
\nabla }_{\mathbf{k}}|u_{n}(\mathbf{k})\rangle $ is the Berry connection,
with $|u_{n}(\mathbf{k})\rangle $ the eigenvector (Bloch wave function) of
the $n$-th band. $\mathcal{C}_{n}=\pm 1$ indicates that the Berry curvature
(flux) $\mathbf{\Omega }_{n}(\mathbf{k})=\mathbf{\nabla }_{\mathbf{k}}\times
\mathbf{\mathcal{A}}_{n}(\mathbf{k})$ on the closed surface $\mathbf{S}$ is
quantized, revealing the synthetic magnetic monopole behavior. The
distribution of Berry curvatures around $W_{1}$ is shown in Fig.~\ref%
{fig:Berry}a (Berry curvatures for different bands and different surface $%
\mathbf{S}$ can be found in Appendix B), yielding $\mathcal{\mathbf{C}}\equiv
\lbrack \mathcal{C}_{1},\mathcal{C}_{2},\mathcal{C}_{3}]=[-1,1,0]$ for the
Weyl point $W_{1}$, and $\mathcal{\mathbf{C}}=[0,1,-1]$ for $W_{2}$. When
both $W_{1}$ and $W_{2}$ are enclosed by $\mathbf{S}$, we have $\mathcal{%
\mathbf{C}}=[-1,2,-1]$. Notice that the Chern numbers remain unchanged even
when the radius of the $\mathbf{S}$ approaches infinity, which explains why
the surface arcs are connected to infinity momenta, indicating that there is
another pair of Weyl points with opposite charges.

\begin{figure}[t]
\includegraphics[width=1.0\linewidth]{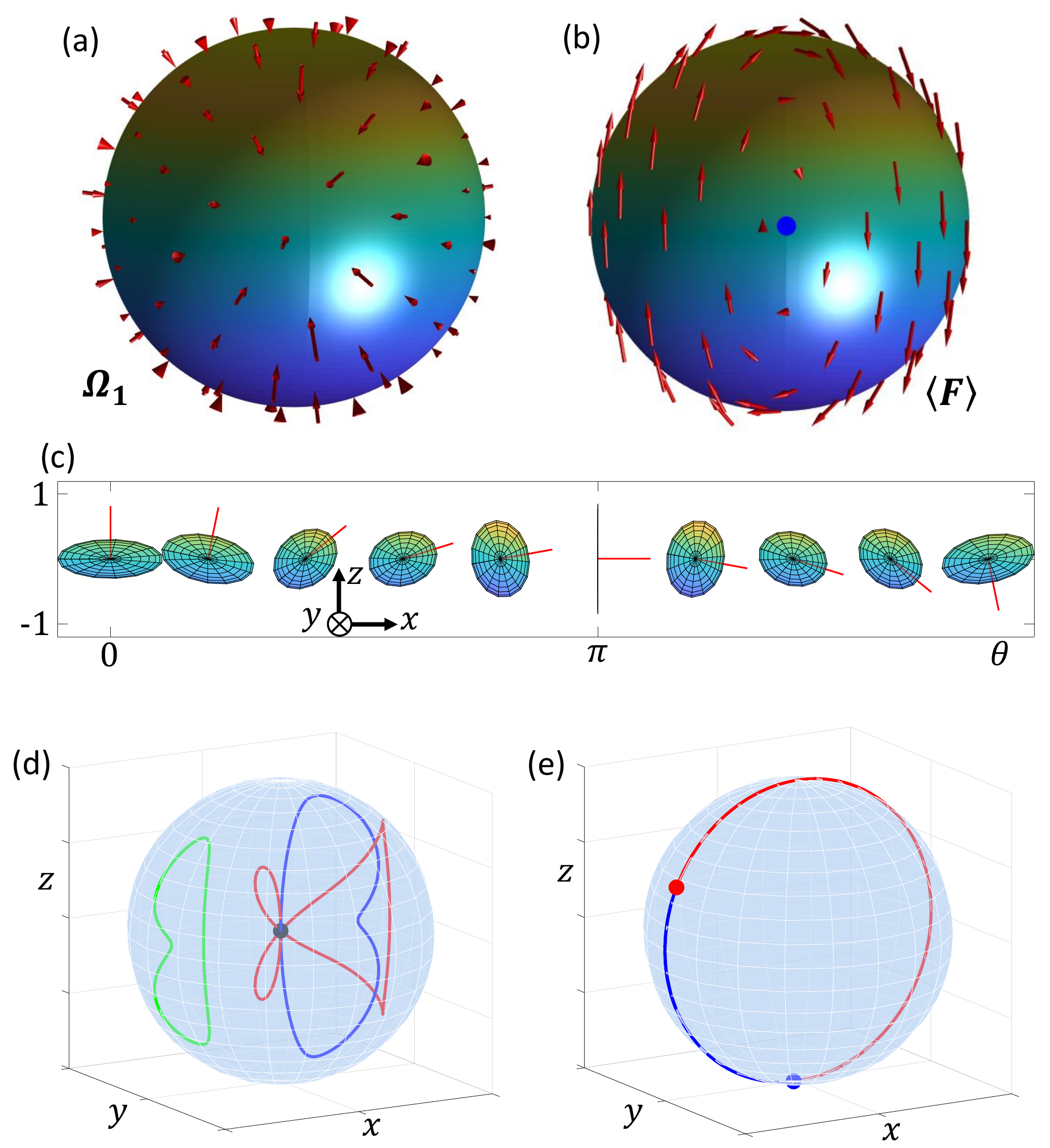}
\caption{(a) and (b) Berry flux and spin vector distributions on the sphere $\mathbf{S}$
enclosing Weyl point $W_{1}$. (c) Spin tensor distribution as $\mathbf{k}$
varies on the loop $\mathcal{L}$. The red lines show the direction of the
ellipsoid's one axis, which is rotated by $\protect\pi $ along the loop. (d) The spin
vector trajectories for the loop $\mathcal{L}$ for all three bands. $\langle
\mathbf{F}\rangle $ would cross the center of the Bloch sphere only for the
two lower non-trivial bands (pink and blue lines). (e) The trajectories of
the two Majorana stars on the Bloch sphere for the loop $\mathcal{L}$ (the
blue and red dots show their initial positions). They exchange their
positions and together give rise to a $\protect\pi $ Berry phase. The radius
of sphere $\mathbf{S}$ is
$r_{S}=1$
and other parameters are the same as in Fig.~\protect\ref{fig:band}.}
\label{fig:Berry}
\end{figure}

For a spin-1/2 system, 
the quantum state is uniquely represented by a point on the Bloch sphere
whose coordinates are given by the expectation value of spin vector $\langle
\mathbf{F}\rangle $. As momentum $\mathbf{k}$ runs over a surface enclosing
a Weyl point in such spin-1/2 system, $\langle \mathbf{F}\rangle $ also
covers the Bloch sphere once, and the Berry flux is given by the solid angle
on the Bloch sphere. Spin-1 (and higher) quantum states are quite different:
first, its quantum state is not uniquely represented by
the spin vector $\langle \mathbf{F}\rangle $; and second, $\langle \mathbf{F}%
\rangle $ is not confined to the surface of the Bloch sphere, and could be
anywhere on or inside the Bloch sphere. For high spins ($\geq 1$), the spin
moments contain both spin vectors and spin tensors. The spin-1 quantum state
can be uniquely represented by the combination of the spin vector $\langle
\mathbf{F}\rangle $ and a rank-2 spin tensor $T$ with elements $%
T_{ij}=\langle \frac{F_{i}F_{j}+F_{j}F_{i}}{2}\rangle -\langle F_{i}\rangle
\langle F_{j}\rangle $, which is geometrically characterized by an ellipsoid.
The ellipsoid is fully determined by its three axes, whose lengths and directions
are given by the eigenvectors and the
square root of the eigenvalues of $T$, respectively~\cite{h2018non,
bharath2018singular,PhysRevB.101.140412}. The topology of the Weyl point in our spin-1 system
should be characterized by the geometries of both the spin vector and tensor
textures, which are fundamentally different from spin-1/2 systems.

An arbitrary spin-1 quantum
state $|\Psi \rangle $ can be characterized by four parameters $F,\phi
_{F},\theta _{F},\phi _{T}$, where $F=|\langle \mathbf{F}\rangle |$ is the
spin-vector length, $\phi _{F},\theta _{F}$ determine the direction of the
spin vector and $\phi _{T}$ gives the relative rotation of the
spin-tensor-ellipsoid with respect to the spin vector~\cite%
{h2018non,bharath2018singular,PhysRevB.101.140412}. This is because, for a given $\langle
\mathbf{F}\rangle $ (i.e., $F,\phi _{F},\theta _{F}$), the size of the
spin-tensor-ellipsoid is also fixed with three axis lengths $\sqrt{1-F^{2}}$%
, $\sqrt{\frac{1\pm \sqrt{1-F^{2}}}{2}}$. Moreover, the axis with length $%
\sqrt{1-F^{2}}$ has the same direction with $\langle \mathbf{F}\rangle $,
and $\phi _{T}$ gives the direction of the other two axes with length
$\sqrt{\frac{1\pm \sqrt{1-F^{2}}}{2}}$%
, which fixes the orientation of the ellipsoid~\cite%
{h2018non,bharath2018singular,PhysRevB.101.140412}. In particular, we have
\begin{equation}
|\Psi (F,\phi _{F},\theta _{F},\phi _{T})\rangle =D(\phi _{F},\theta _{F},\phi _{T})\left[
\begin{array}{c}
\sqrt{\frac{1+F}{2}} \\
0 \\
\sqrt{\frac{1-F}{2}}%
\end{array}%
\right]   \label{eq:state_Tensor_S}
\end{equation}
with $D(\phi _{F},\theta _{F},\phi _{T})=e^{-iF_{z}\phi
_{F}}e^{-iF_{y}\theta _{F}}e^{-iF_{z}\phi _{T}}$.
Let’s consider an infinitesimal sphere $\mathbf{S}$ enclosing the Weyl point,
the state of the third far gapped band remains unchanged on the whole sphere since the sphere is infinitesimal.
We can denote the third-band state at the Weyl point as $|u^0_3\rangle=|\Psi (F,\phi _{F},\theta _{F},\phi _{T})\rangle$.
We first consider $F\neq0$, there exits one and only one state $|u^0_2\rangle=|\Psi (F,\phi _{F},\theta _{F}+\pi,\phi _{T}+\pi)\rangle$
satisfying $\langle u^0_3|u^0_2\rangle=0$ and $\langle u^0_3|\mathbf{F}|u^0_3\rangle=-\langle u^0_2|\mathbf{F}|u^0_2\rangle$.
Notice that $\mathbf{F}$ is traceless $\text{tr}(\mathbf{F})=0$,
therefore, there exist one and only one state satisfying $\langle u^0_3|u^0_1\rangle=0$ and
$\langle u^0_1|\mathbf{F}|u^0_1\rangle=0$.
Naturally, we also have $\langle u^0_2|u^0_1\rangle=0$.
The Weyl point is characterized by an effective spin-1/2 system spanned by states $|u^0_1\rangle$ and $|u^0_2\rangle$.
As the momentum changes over the sphere $\mathbf{S}$,
the eigenstate of the two Weyl bands changes on the Bloch sphere spanned by $|u^0_1\rangle$ and $|u^0_2\rangle$.
The Chern number counts the times the eigenstate covers the Bloch sphere. $|u^0_1\rangle$
is the only state on the Bloch sphere that gives vanishing $\langle \mathbf{F}\rangle$,
therefore, the Chern number is odd (even) if and only if there are odd (even) numbers vanishing points of the spin vector.
For the special case with $F=0$,
it can be shown that $\langle \mathbf{F}\rangle$ can vanish on a loop (a great circle coinciding with the
prime meridian) instead of a point on the Bloch sphere,
and thus the Chern number is odd (even) if and only if there are odd (even) numbers of vanishing loops of the spin vector.
The spin vectors form a vortex (change their sign)
around the vanishing point (across the vanishing loop).
Therefore, the spin-vector vanishing points and loops are topological structures that
can only change abruptly (e.g., at a gap closing), and they will
remain unchanged under smooth deformations (e.g., enlarging the sphere $\mathbf{S}$).

In Fig.~\ref{fig:Berry}b, we show the spin vector distribution $\langle
\mathbf{F}\rangle $ (calculated for the lowest band) around $W_{1}$ (spin
textures for different bands and different surfaces $\mathbf{S}$ can be found
in Appendix B). We see that the spin vector may vanish at one certain point (blue
dot), around which spin vortex emerges.
Though the Chern number of the Weyl point
can be obtained by the number of spin vortices,
the Berry phase along a loop (i.e., Berry
flux on a surface enclosed by the loop) in the momentum space (or other
parameter space) is given by the generalized solid angles involving
contributions from both spin vectors and tensors on or inside the Bloch
sphere (see Appendix C).

To illustrate how the spin tensor is
distributed around $W_{1}$, we consider a loop on $\mathbf{S}$, and study
how the ellipsoid rotates along it. Fig.~\ref{fig:Berry}c shows the spin
tensor (with view direction along $y$ axis) for the first band along the loop $\mathcal{L}$: $k_{x}=r_{S}\cos
\theta $, $k_{y}=r_{S}\sin \theta $, $k_{z}=0$ with $r_{S}$ the radius and $%
\theta $ varies from 0 to $2\pi $ (i.e., the equator of $\mathbf{S}$).
As $\theta $ increases from 0 to $2\pi $, the ellipsoid is reduced to a 2D
disk at $\theta =\pi $, where the spin vector crosses the center of the
Bloch sphere as shown in Fig.~\ref{fig:Berry}d ($\langle \mathbf{F}\rangle
$ vanishes and changes the sign).
Along the loop $\mathcal{L}$, the spin vector is confined
in the $F_{x}$-$F_{z}$ plane, which gives rise to zero solid angle. Beside
the size oscillation, the orientation of the spin tensor ellipsoid rotates
around the $y$-axis by $\pi $, which correspond to a $\pi $ Berry phase
along $\mathcal{L}$ due to the fact that the Weyl point reduces to a Dirac
point in the $k_{z}=0$ plane. Along the loop $\mathcal{L}$,
one of the ellipsoid's axes is fixed along
$y$ direction, and its length is around 0.85 which changes slightly with $\theta$.
Similar spin tensor rotation can be obtained
for the second band. However, the spin vector crosses the center of the
Bloch sphere three times on the loop $\mathcal{L}$, leading to three
spin-vector vortices on $\mathbf{S}$.

The nontrivial topology of the Weyl points can also be captured by the
trajectories of two Majorana stars (an unordered pair of points on the Bloch
sphere)~\cite{PhysRevLett.113.240403}. The Berry flux is given by the
correlated solid angle of the two Majorana stars (see Appendix D). 
For the loop $\mathcal{L}$ considered in Fig.~\ref{fig:Berry}c, we find that
the Majorana stars are confined in the $y=0$ plane on the Bloch sphere. As $%
\theta $ increases, similar to the spin-tensor ellipsoid, the Majorana stars
also rotate with respect to $y$-axis. Instead of going back to their
originate positions after one circle, two Majorana stars exchange as shown
in Fig.~\ref{fig:Berry}e,
leading to a solid angle $\pi $.

\section{Implementation and detection}
 Our scheme does not
rely on atomic hyperfine level structure, and is applicable to both alkaline
atoms (e.g., Lithium, potassium) and alkaline-earth(-like) atoms (e.g.,
strontium, ytterbium)~\cite{PhysRevLett.117.220401, mancini2015observation,
kolkowitz2017spin, bromley2018dynamics}. The triple-well superlattice could
be realized by a superposition of two lattice potentials with one of them
having a tripled period,
\begin{equation}
V(z)=V_{1}\cos ^{2}(k_{L}z)+V_{2}\cos ^{2}(k_{L}z/3+\phi _{L}).
\end{equation}%
Using optical frequency tripling~\cite{pfister1997continuous,
fedotov1991highly}, such two lattice potential can be obtained with tunable
relative phase $\phi _{L}$, similar as the double well superlattice based on
the optical frequency doubling in recent experiments~\cite%
{lohse2016thouless, aidelsburger2015measuring, li2016spin, li2017stripe}.
Alternatively, it can also be realized using lasers with the same
wavelength, while the long-period lattice is formed by two beams
intersecting with an angle $\theta =2\arcsin \frac{1}{3}$. By choosing
proper lattice strengths $V_{1}$, $V_{2}$ and the relative phase $\phi _{L}$%
, the detunings between different sites in a unit cell is tuned to be much
larger than the bare nearest neighbor tunneling. The tunnelings can be
restored using resonant Raman couplings, as demonstrated by recent
experiments in the study of gauge field and supersolidity.

The linear dispersion of the Weyl point can be detected using
momentum-resolved radio-frequency (rf) spectroscopy~\cite{zhang2014fermi},
which has been widely used to study low-energy excitation spectrum and
quasiparticles in superfluids and superconductors. Based on energy and
momentum conservation, the Weyl point dispersion can be extracted from the
time-of-flight absorption image after the rf pulse. In general, direct
measurement of non-trivial Berry curvatures and spin textures of Weyl points
is very challenging, and simple schemes for probing Weyl-point topology are
still elusive.
Here we propose that the detection can be realized by the momentum-resolved
Rabi spectroscopy~\cite{wall2016synthetic} with simple pulse sequences.
Surprisingly, the simplification comes from the presence of the third band
near the Weyl point for our spin-1 system. First, the system is initialized
into the pseudo-spin state $|s\rangle $, then the Raman lasers are turned
on. By simply measuring the evolution of atom population on state $%
|s^{\prime }\rangle $ at each $\mathbf{k}$, the Bloch wave-function (and
thereby the Berry curvatures and spin textures) near the Weyl points can be
extracted. There is no need to measure the population on different
basis as required for spin-1/2 systems. This is because, beside two
non-trivial bands, there is a far detuned trivial band near the Weyl point,
which can serve as a reference band, allowing us to determine both
amplitudes and phases of the Bloch functions for two non-trivial bands.
In realistic experiments, the population of each spin state $|s\rangle
$ at each $\mathbf{k}$ can be measured using a pseudospin Stern-Gerlach
effect followed by the time-of-flight imaging~\cite{li2016spin, li2017stripe}%
.

It has been demonstrated that, for a
spin-1/2 system, the Bloch wave function, which directly determines the
Berry curvatures and spin textures, can be extracted from the
momentum-resolved Rabi spectroscopy realized by proper choice of laser pulse
sequences~\cite{PhysRevA.100.063630}.
Surprisingly, for our spin-1 system, the presence of a third band
would greatly simplify the pulse sequence. The Bloch wavefunction of the $n$%
-th band with energy $E_{n}(\mathbf{k})$ is
\begin{equation}
|u_{n}(\mathbf{k})\rangle =\sum_{s}U_{n,s}(\mathbf{k})|\mathbf{k},s\rangle ,
\end{equation}%
with $U_{n,s}(\mathbf{k})$ the element of the unitary matrix $U$. Consider
an initial state $|\Psi (0)\rangle =|\mathbf{k},s\rangle $, the Hamiltonian
would induce a Rabi oscillation and give a final state at time $\tau $
\begin{equation}
|\Psi (\tau )\rangle =\sum_{n}e^{-iE_{n}(\mathbf{k})\tau }U_{n,s}^{\ast }(%
\mathbf{k})|u_{n}(\mathbf{k})\rangle .
\end{equation}%
In the following, we prove that the Bloch wave function can be obtained by
simply measuring the final state in the spin basis $\{|\mathbf{k},s\rangle
\} $ with $s=\uparrow ,0,\downarrow $. Thanks to the presence of the third
band, the detecting scheme is simpler comparing with the spin-1/2 system
(where measurements in various bases and thus additional precisely controlled
pulses are required)~\cite{PhysRevA.100.063630}.

The population on state $|\mathbf{k},s^{\prime }\rangle $ of the final state
is
\begin{equation}
P_{s,s^{\prime }}(\mathbf{k},\tau )=\left\vert \sum_{n}e^{-iE_{n}(\mathbf{k}%
)\tau }U_{n,s}^{\ast }(\mathbf{k})U_{n,s^{\prime }}(\mathbf{k})\right\vert
^{2}.
\end{equation}%
We define the averaged population $\bar{P}_{s,s^{\prime }}$ as
\begin{equation}
\bar{P}_{s,s^{\prime }}=\frac{P_{s,s^{\prime }}+P_{s^{\prime },s}}{2}.
\end{equation}%
Use the Fourier analysis in the time domain $\bar{P}_{s,s^{\prime }}(\mathbf{%
k},\omega )=\int d\tau \bar{P}_{s,s^{\prime }}(\mathbf{k},\tau )\cos (\omega
\tau )$, we obtain
\begin{eqnarray}
\bar{P}_{s,s^{\prime }}(\mathbf{k},\omega )=\sum_{n<n^{\prime
}}\bigg[|U_{n,s}^{\ast }U_{n,s^{\prime }}U_{n^{\prime },s^{\prime }}^{\ast
}U_{n^{\prime },s} | \nonumber \\
\times\cos (\phi _{n;s,s^{\prime }}+\phi _{n^{\prime
};s^{\prime },s})
\delta (E_{n^{\prime }}-E_{n}-\omega )\bigg],
\end{eqnarray}%
where relative phase $\phi _{n;s,s^{\prime }}=\phi _{n;s^{\prime }}-\phi
_{n;s}$ with $\phi _{n;s}=\arg [U_{n,s}]$. For $s^{\prime }=s$, we can
easily obtain the amplitude of the matrix elements $|U_{n,s}|$ based on $%
\bar{P}_{s,s^{\prime }}(\mathbf{k},\omega )$ and the unitary property of
matrix $U$. Extracting the phase information is, however, a little bit
tricky. For Weyl points in a spin-1/2 system, it is impossible to determine
the phase $\phi _{n;s,s^{\prime }}$ from $\bar{P}_{s,s^{\prime }}(\mathbf{k}%
,\omega )$ since both $\phi _{n;s,s^{\prime }}$ and $\phi _{n^{\prime
};s^{\prime },s}$ changes rapidly near the Weyl point and one can only
obtain their summation $\phi _{n;s,s^{\prime }}+\phi _{n^{\prime };s^{\prime
},s}$ (not to mention that this summation usually vanishes). However, for a
spin-1 system, the third band can serve as a reference which allows the
determination of the phases for the other two bands.

To show how our detecting scheme works, we focus our discussion on Weyl
point $W_{1}$ in the following. In the vicinity of Weyl point $W_{1}$, the
Bloch wavefunctions possess non-trivial topology due to the degeneracy for
two lower bands, but are trivial and almost unchanged for the highest band.
Near the frequency $\omega =E_{3}-E_{n}$, we have
\begin{equation}
\bar{P}_{s,s^{\prime }}\propto |U_{n,s}^{\ast }U_{n,s^{\prime
}}U_{3,s^{\prime }}^{\ast }U_{3,s}|\cos (\phi _{n;s,s^{\prime }}-\phi
_{3;s,s^{\prime }}),
\end{equation}%
where $\phi _{3;s,s^{\prime }}$ is a constant near the Weyl point, and can
be set to zero by absorbing it to the definition of $|\mathbf{k},s\rangle $.
Therefore we obtain the relative phase $\phi _{n;s,s^{\prime }}$ (with $%
n=1,2 $) for the two non-trivial bands through measuring $\bar{P}%
_{s,s^{\prime }}$. In fact, even the phase $\phi _{3;s,s^{\prime }}$ is not
a constant, the topologies of the Bloch functions are not affected by
absorbing $\phi _{3;s,s^{\prime }}$ into the definition of $|\mathbf{k}%
,s\rangle $, as long as $\phi _{3;s,s^{\prime }}$ is a non-singular and
smooth function near the Weyl point $W_{1}$. $|U_{n,s}|$ and $\phi
_{n;s,s^{\prime }}$ can be uniquely determined in a way such that the Bloch
wave function is smooth.

The measured relative phase, which is used to extract the Bloch wave
function, is $\phi _{n;s,s^{\prime }}^{M}=\phi _{n;s,s^{\prime }}-\phi
_{3;s,s^{\prime }}$. As a result, the measured Bloch wave function $%
|u_{n}^{M}(\mathbf{k})\rangle $ and the true Bloch wave function $|u_{n}(%
\mathbf{k})\rangle $ are related by a unitary transformation $|u_{n}^{M}(%
\mathbf{k})\rangle =e^{-i\hat{\Phi}}|u_{n}(\mathbf{k})\rangle $, with $\hat{%
\Phi}=\text{diag}\{\phi _{3;\uparrow },\phi _{3;0},\phi _{3;\downarrow }\}$.
The measured Chern number using $|u_{n}^{M}(\mathbf{k})\rangle $ is
\begin{equation}
\mathcal{C}_{n}^{M}=\frac{1}{2\pi }\oint_{\mathbf{S}}\left[ \mathbf{\Omega }%
_{n}+\mathbf{\nabla }_{\mathbf{k}}\times \langle u_{n}(\mathbf{k})|\hat{%
\mathbf{\chi }}|u_{n}(\mathbf{k})\rangle \right] \cdot d\mathbf{S},
\end{equation}%
with $\hat{\mathbf{\chi }}=\mathbf{\nabla }_{\mathbf{k}}\hat{\Phi}$. In the
very vicinity of the Weyl point $W_{1}$, $\hat{\Phi}$ is a constant diagonal
matrix, and the second term in the square brackets of the above equation
vanishes. Therefore the measured Berry curvature and Chern number are the
same as their true values. Far away from the Weyl point, $\hat{\Phi}$
becomes $\mathbf{k}$ dependent, and the measured Berry curvature may have
small derivations from the true value, however, the measured Chern number is
unaffected as long as $e^{-i\hat{\Phi}}$ is non-singular and smooth, which
holds for our case when $\mathbf{S}$ only encloses one Weyl point $W_{1}$.

\begin{figure}[t]
\includegraphics[width=1.0\linewidth]{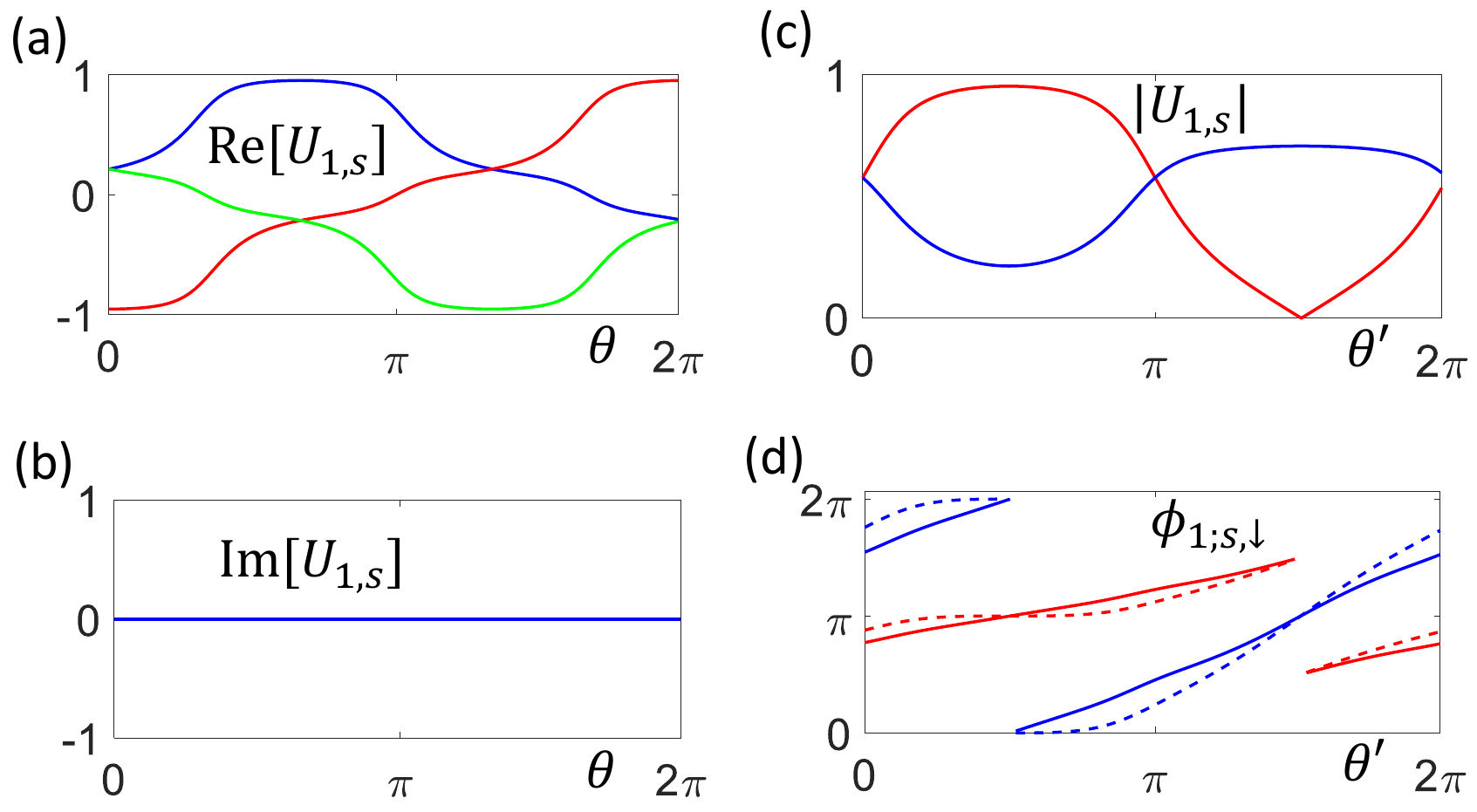}
\caption{(a) and (b) The real and imaginary parts of $U_{1,s}$ on loop $%
\mathcal{L}$: $k_{x}=\cos \protect\theta $, $k_{y}=\sin \protect\theta $, $%
k_{z}=0$ around $W_{1}$, with blue, red and green lines corresponding to the
spin states $s=\uparrow $, $s=0$ and $s=\downarrow $, respectively. (c) and
(d) The amplitude and phase of $U_{1,s}$ on loop $\mathcal{L}^{\prime }$: $%
k_{z}=\cos \protect\theta $, $k_{x}=\sin \protect\theta $, $k_{y}=0$, with
blue and red lines corresponding to the spin states $s=\uparrow $ and $s=0$,
respectively. The amplitude for $s=\downarrow $ is the same as that for $%
s=\uparrow $, and the phase is measured with respect to spin state $%
s=\downarrow $. In (d), dashed and solid lines correspond to the measured
and true values, respectively. Other parameters are the same as in Fig.~\ref{fig:Berry}.}
\label{fig:Rabi}
\end{figure}

In Fig.~\ref{fig:Rabi}, we show numerical results for the phases extracted
from $\bar{P}_{s,s^{\prime }}$ and their true values obtained directly from
the Hamiltonian on the two loops $\mathcal{L}$ and $\mathcal{L}^{\prime }$.
For the loop $\mathcal{L}$, we always have $\phi _{n;s,s^{\prime }}=0,\pi $,
so the Bloch wavefunction can be extracted solely from $|U_{n,s}|$, while
the phase can be determined simply by the continuous properties. For the
loop $\mathcal{L}^{\prime }$ with a large radius $r_{S}=1$, we see small
derivations of the measured relative phases from their true values.

In realistic experiments, the initialization is realized by first tuning the
lattice potential such that the $s$-sites have the lowest energy in each
unit cell, loading atoms to the pseudo-spin state $|s\rangle $, and then
adiabatically tuning the potential to the desired superlattices. Next, we
can turn on the Raman lasers and let the system evolve with an interval $%
\tau $. The population of the final state on $s^{\prime }$-sites at each $%
\mathbf{k}$ (i.e., $P_{s,s^{\prime }}(\mathbf{k},\tau )$) can be measured
using a pseudospin Stern-Gerlach effect followed by the time-of-flight
imaging~\cite{li2016spin, li2017stripe}.

\section{Conclusion}
In summary, we propose a simple scheme
to realize robust Weyl points and probe their topology, using a 1D
triple-well superlattice with transverse 2D SOC generated by three Raman
lasers. The robustness against system parameters such as laser intensities,
phases, polarizations and incident angles makes our scheme very flexible,
and any fine-tuning or phase-locking techniques are not required. Moreover,
we find that the spin-1 Weyl point shows very interesting and topologically
non-trivial spin (vector and tensor) textures that have fundamental
differences from spin-1/2 systems. Thanks to the three-band structure, these
non-trivial topologies can be detected using very simple pulse sequences. A
straightforward generalization of our scheme is to consider higher-order
degeneracies (e.g., three- or four-fold)~\cite{bradlyn2016beyond,
lv2017observation,PhysRevLett.120.240401,PhysRevA.98.013627} using even higher spins, which may be realized by using a
superlattice with more sites in each unit-cell or by including atomic
hyperfine states. Our scheme provides a simple yet powerful platform for
exploring Weyl physics and related high-dimensional topological phenomena
with ultracold atoms.

\begin{acknowledgments}
\textbf{Acknowledgements}: This work is supported by AFOSR
(FA9550-16-1-0387), NSF (PHY-1806227), and ARO (W911NF-17-1-0128).
\end{acknowledgments}




\section*{Appendix}

\subsection{A. The Weyl point solution}

\begin{figure}[tb]
\includegraphics[width=1.0\linewidth]{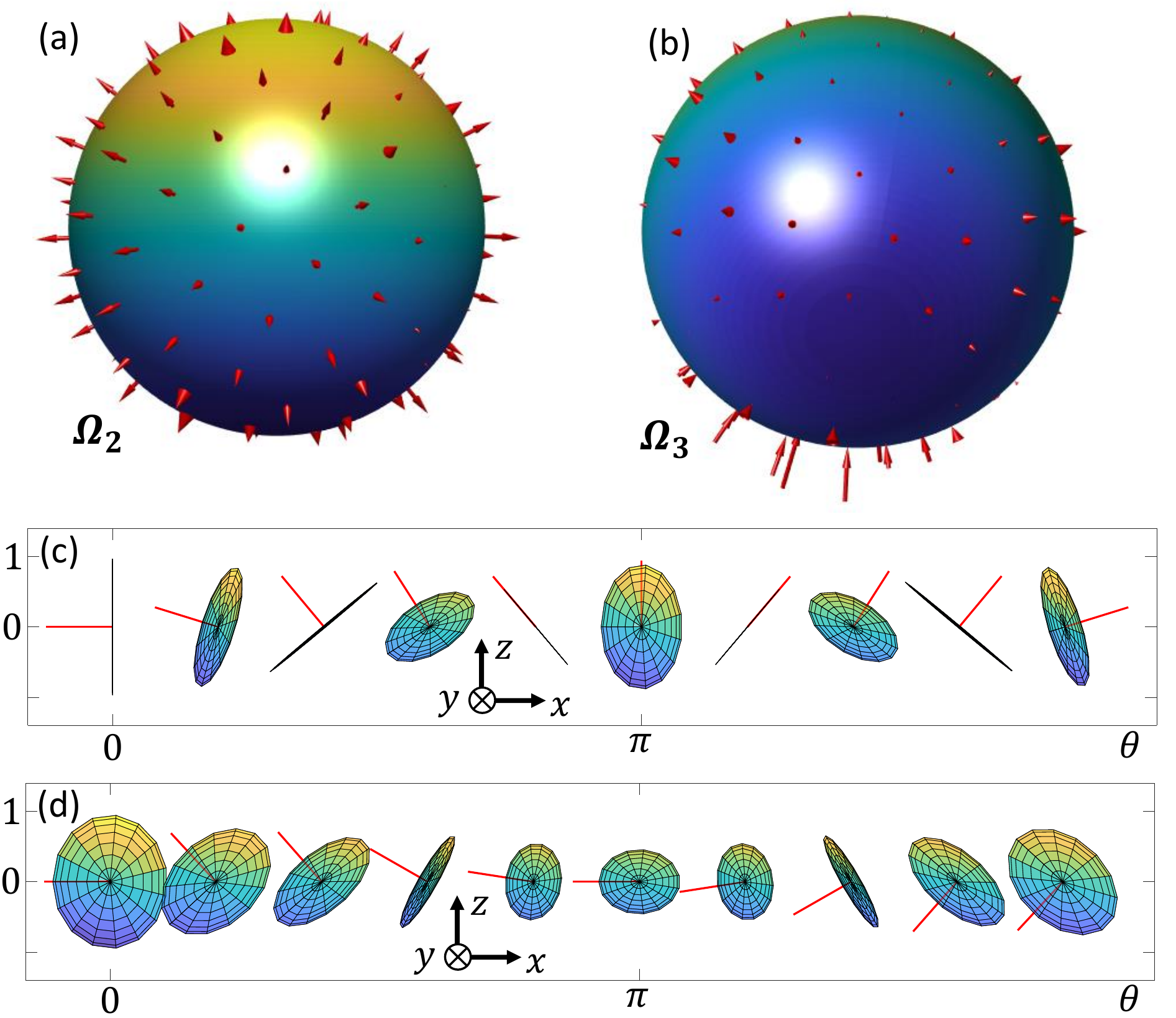}
\caption{(a) and (b) The Berry flux distributions on the surface enclosing
Weyl point $W_{1}$ for the second and third bands, respectively. (c) and (d)
Spin tensor distributions as $\mathbf{k}$ varies on the loop $\mathcal{L}$
for the second and third band, respectively. The red lines show the
orientation of the ellipsoids, which are rotated by $\protect\pi $ (0) for
the second (third) band. Other parameters are the same as in Fig.~\ref{fig:Berry}.}
\label{fig:band23}
\end{figure}

In the basis $\{\left|\uparrow \right\rangle ,|0\rangle ,\left|\downarrow \right\rangle \}$, the
momentum space Hamiltonian is
\begin{eqnarray}
H_{\mathbf{k}}&=&\left[
\begin{array}{ccc}
\bar{\delta}_{\uparrow }-\frac{\mathbf{k}_{\bot }\cdot \mathbf{K}_{\bot ,\uparrow
}}{m} & J_{1} & J_{3}e^{ik_{z}} \\
J_{1} & \bar{\delta}_{0}-\frac{\mathbf{k}_{\bot }\cdot \mathbf{K}_{\bot ,0}}{m} & J_{2} \\
J_{3}e^{-ik_{z}} & J_{2} & \bar{\delta}_{\downarrow }-\frac{\mathbf{k}_{\bot
}\cdot \mathbf{K}_{\bot ,\downarrow }}{m}%
\end{array}%
\right] \nonumber \\
&&+\frac{\mathbf{k}_{\bot }^2}{2m}
\end{eqnarray}%
with $\bar{\delta}_{s}=\delta_{s}+\frac{\mathbf{K}_{\bot,s }^2}{2m}$.
We can redefine the Fermi energy $E_{F}$ as the zero energy point, and
rewrite the Hamiltonian as
\begin{eqnarray}
H_{\mathbf{k}}&=&\left[
\begin{array}{ccc}
\bar{\delta}_{\uparrow }-\frac{\mathbf{k}_{\bot }\cdot \mathbf{K}_{\bot ,\uparrow
}}{m} & J_{1} & J_{3}e^{ik_{z}} \\
J_{1} & \bar{\delta}_{0}-\frac{\mathbf{k}_{\bot }\cdot \mathbf{K}_{\bot ,0}}{m} & J_{2} \\
J_{3}e^{-ik_{z}} & J_{2} & \bar{\delta}_{\downarrow }-\frac{\mathbf{k}_{\bot
}\cdot \mathbf{K}_{\bot ,\downarrow }}{m}%
\end{array}%
\right] \nonumber \\
&&+\frac{\mathbf{k}_{\bot }^2}{2m}-E_F
\end{eqnarray}%
The Weyl point corresponds to a two-fold degeneracy with zero energy, which
requires that there exist a Fermi energy $E_{F}$ and a momentum $\mathbf{k}%
^{W}$ such that the above Hamiltonian is a rank-1 matrix. At the Weyl
points, $E_{F}$ and $\mathbf{k}^{W}$ correspond to the solutions of the
following equations
\begin{equation}
\left\{
\begin{array}{c}
\frac{\bar{\delta}_{\uparrow }-\frac{\mathbf{k}_{\bot }\cdot \mathbf{K}_{\bot
,\uparrow }}{m}+E_{\bot }-E_{F}}{J_{1}}=\frac{J_{1}}{-\frac{\mathbf{k}_{\bot }\cdot
\mathbf{K}_{\bot ,0}}{m}+E_{\bot }-E_{F}}=\frac{J_{3}e^{ik_{z}}}{J_{2}} \\
\frac{\bar{\delta}_{\uparrow }-\frac{\mathbf{k}_{\bot }\cdot \mathbf{K}_{\bot
,\uparrow }}{m}+E_{\bot }-E_{F}}{J_{3}e^{-ik_{z}}}=\frac{J_{1}}{J_{2}}=\frac{%
J_{3}e^{ik_{z}}}{\bar{\delta}_{\downarrow }-\frac{\mathbf{k}_{\bot }\cdot \mathbf{%
K}_{\bot ,\downarrow }}{m}+E_{\bot }-E_{F}}.%
\end{array}%
\right.
\end{equation}%
We have set $\bar{\delta}_{0}=0$ without loss of generality.
Therefore, we have $k_{z}^{W}=0$ or $\pi $, and $\mathbf{k}_{\bot }^{W}$ is the solution of the equations
\begin{equation}
\left\{
\begin{array}{rcl}
\frac{J_{1}J_{3}e^{ik_{z}^{W}}}{J_{2}}-\frac{J_{1}J_{2}}{J_{3}e^{ik_{z}^{W}}}
& = & \frac{\mathbf{k}_{\bot }\cdot (\mathbf{K}_{\bot ,0}-\mathbf{K}_{\bot
,\uparrow })}{m}+\bar{\delta}_{\uparrow } \\
\frac{J_{2}J_{3}e^{ik_{z}^{W}}}{J_{1}}-\frac{J_{1}J_{2}}{J_{3}e^{ik_{z}^{W}}}
& = & \frac{\mathbf{k}_{\bot }\cdot (\mathbf{K}_{\bot ,0}-\mathbf{K}_{\bot
,\downarrow })}{m}+\bar{\delta}_{\downarrow } %
\end{array}%
\right.
\end{equation}%
We always have solutions as long as $J_{1,2,3}$ are nonzero and $\mathbf{K}%
_{\bot ,0}-\mathbf{K}_{\bot ,\uparrow }$ is not parallel with $\mathbf{K}%
_{\bot ,0}-\mathbf{K}_{\bot ,\downarrow }$ (i.e., the three points $\mathbf{K%
}_{\bot ,s}=(K_{x,s},K_{y,s})$ are not collinear), which share the same
spirit as the recent study of Dirac degeneracy with 2D spin-orbit coupling~%
\cite{meng2016experimental, huang2016experimental}.
The fermi energy is given by $E_{F} =  \frac{(\mathbf{k}_{\bot }^{W})^2}{2m}-\frac{J_{1}J_{2}}{J_{3}e^{ik_{z}^{W}}}-\frac{\mathbf{k}%
_{\bot }^{W}\cdot \mathbf{K}_{\bot ,0}}{m}$.
We would like to point out that,
for pseudo-spin states represented by the atomic hyperfine levels as in~%
\cite{meng2016experimental, huang2016experimental} (where the 2D Dirac
degeneracy is sensitive to the Raman-laser polarizations), it is not
easy to generalize the 2D Dirac degeneracy to 3D Weyl degeneracy.

\begin{figure}[tb]
\includegraphics[width=1.0\linewidth]{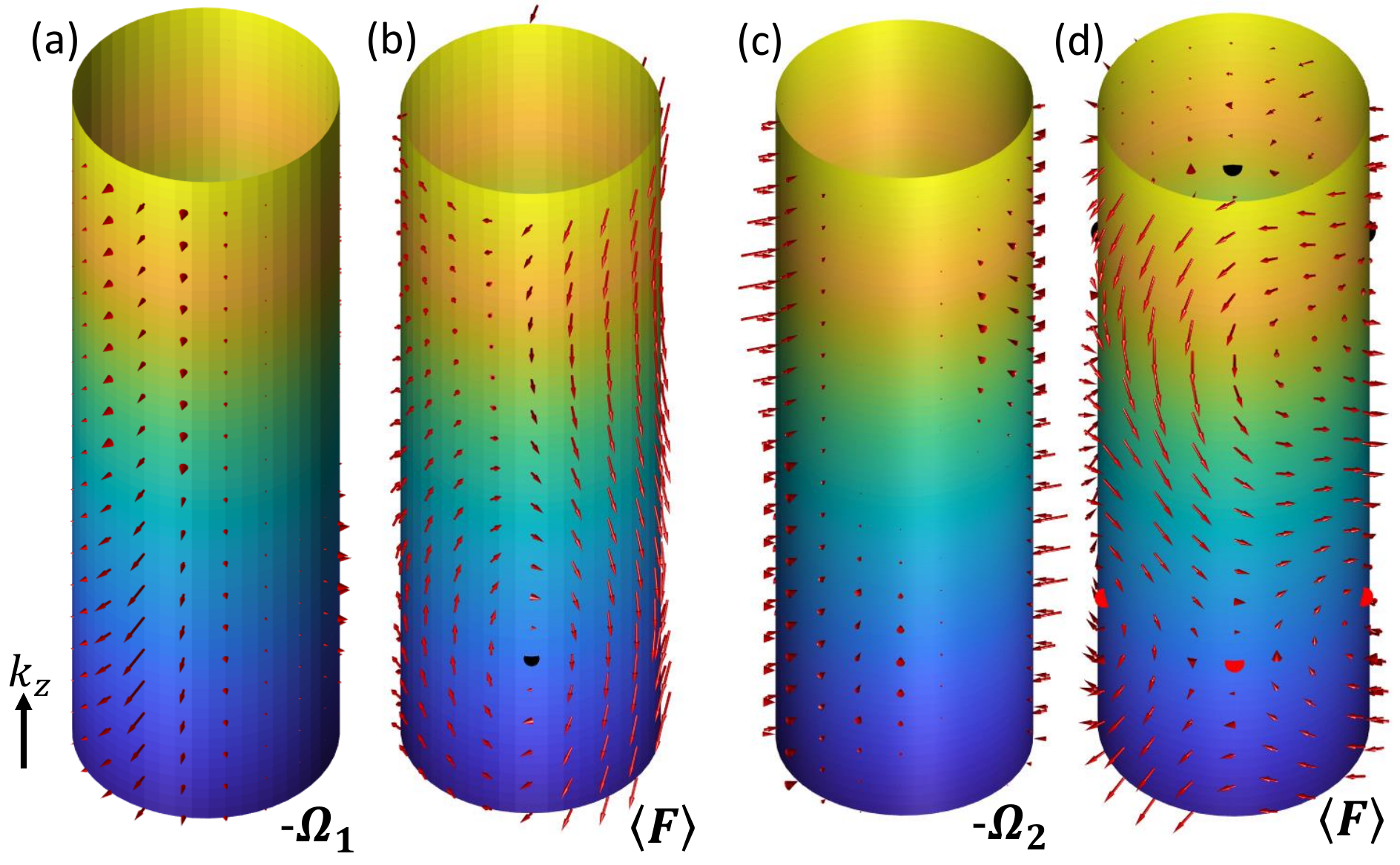}
\caption{(a) and (b) The Berry flux and spin vector distributions of the
first band on the surface enclosing both Weyl points $W_{1}$ and $W_{2}$. A
spin vortex (black dot) is located at $k_{z}=0$. (c) and (d) The Berry flux
and Spin vector distributions of the second band on the surface enclosing
both Weyl points $W_{1}$ and $W_{2}$. Six spin vortices (red and black dots)
are located at $k_{z}=0$ and $k_{z}=\protect\pi $. The radius of the
cylinder is $1$, other parameters are the same as in Fig.~\ref{fig:Berry}. }
\label{fig:cylinder}
\end{figure}

\subsection{B. Berry flux and spin textures on different surfaces}

As we discussed in the main text, we have $\mathcal{\mathbf{C}}\equiv
\lbrack \mathcal{C}_{1},\mathcal{C}_{2},\mathcal{C}_{3}]=[-1,1,0]$ for the
Weyl point $W_{1}$, and $\mathcal{\mathbf{C}}=[0,1,-1]$ for $W_{2}$. When
both $W_{1}$ and $W_{2}$ are enclosed by $\mathbf{S}$, we have $\mathcal{%
\mathbf{C}}=[-1,2,-1]$. In Fig.~\ref{fig:Berry}a in the main text, we plot the Berry
curvature distribution of the first band around $W_{1}$. Figs.~\ref%
{fig:band23}a and ~\ref{fig:band23}b show the corresponding Berry
curvatures for the other two bands, we see that the total flux for the
second (third) band is quantized to $1$ (0). Such non-trivial topology can
also be characterized by the spin (vector and tensor) textures. For the
first band, the spin tensor is rotated by $\pi $ on the loop $\mathcal{L}$: $%
k_{x}=r_{S}\cos \theta $, $k_{y}=r_{S}\sin \theta $, $k_{z}=0$ with $\theta
\in \lbrack 0,2\pi )$, and the spin vector crosses the center of the Bloch
sphere once (as shown in Figs.~\ref{fig:Berry}c and \ref{fig:Berry}d in the main text), leading to the
generalized solid angle $\gamma_F=0$, $\gamma_T=\pi$. Similarly, for the
second band which is also non-trivial around $W_{1}$, the spin tensor is
also rotated by $\pi $ on the loop $\mathcal{L}$ [see Fig.~\ref{fig:band23}%
c], while the spin vector crosses the center of the Bloch sphere three
times (see Fig.~\ref{fig:Berry}d in the main text), leading to the generalized solid
angle $\gamma_F=0$, $\gamma_T=\pi$. For the trivial third band around $W_{1}$%
, neither $\pi $-rotation for the spin tensor nor Bloch center crossing for
the spin vector would exist [see Fig.~\ref{fig:band23}d], leading to the
generalized solid angle $\gamma_F=0$, $\gamma_T=0$. We may also consider a
different loop $\mathcal{L}^{\prime }$: $k_{z}=r_{S}\cos \theta $, $%
k_{x}=r_{S}\sin \theta $, $k_{y}=0$ with $\theta \in \lbrack 0,2\pi )$
around $W_{1}$, and the spin textures are quite similar with the loop $%
\mathcal{L}$.

When both $W_1$ and $W_2$ are enclosed by the momentum surface $\mathbf{S}$,
the Berry flux is quantized as $\mathcal{\mathbf{C}}=[-1,2,-1]$ [as shown in
Figs.~\ref{fig:cylinder}a and ~\ref{fig:cylinder}c for the first two
bands where $\mathbf{S}$ is a cylinder covering the whole Brillouin zone in $%
k_z$]. The spin vector distributions are shown in Figs.~\ref{fig:cylinder}%
b and ~\ref{fig:cylinder}d for the first two bands. We see that there is
a vortex at $k_z=0$ for the first band, and six vortices (three at $k_z=0$
and the other three at $k_z=\pi$) for the second band. While the Berry flux
and spin distributions for the third band are similar with that for the
first band, except that the vortex is located at $k_z=\pi$.

\subsection{C. Geometric representation of Berry flux}

Consider the parameter $\tau $-dependent Hamiltonian $H(\tau )$. For an
arbitrary loop in the parameter space $\tau \in \lbrack \tau _{\text{i}%
},\tau _{\text{f}}]$ with the Hamiltonian satisfying $H(\tau _{\text{i}%
})=H(\tau _{\text{f}})$, the corresponding Berry phase of a given gapped
eigenstate is
\begin{equation}
\gamma =i\int d\tau \langle \Psi (\tau )|\partial _{\tau }|\Psi (\tau
)\rangle +\gamma _{\text{f,i}},  \label{eq:Berry_S}
\end{equation}%
where $|\Psi (\tau )\rangle $, a smooth function of $\tau $, is the
eigenstate of $H(\tau )$, and $\gamma _{\text{f,i}}$ is the gauge difference
between two ends of the loop that is given by $|\Psi (\tau _{\text{f}%
})\rangle =e^{i\gamma _{\text{f,i}}}|\Psi (\tau _{\text{i}})\rangle $. We
choose four parameters $F(\tau ),\phi _{F}(\tau ),\theta _{F}(\tau ),\phi
_{T}(\tau )$ to ensure a smooth wavefunction $|\Psi (\tau )\rangle =|\Psi
\lbrack F(\tau ),\phi _{F}(\tau ),\theta _{F}(\tau ),\phi _{T}(\tau
)]\rangle $. Substitute Eq.~(\ref{eq:state_Tensor_S}) into Eq.~(\ref{eq:Berry_S})%
, we obtain
\begin{equation}
\gamma =\int [Fd\phi _{T}+F\cos (\theta _{F})d\phi _{F}]+\gamma _{\text{f,i}}
\label{eq:Berry2_S}
\end{equation}%
with $\gamma _{\text{f,i}}=[\phi _{F}(\tau _{\text{i}})-\phi _{F}(\tau _{%
\text{f}})]+[\phi _{T}(\tau _{\text{i}})-\phi _{T}(\tau _{\text{f}})]$. We
now define the generalized solid angle on the loop for the spin vector and
tensor as $\gamma _{F}$ and $\gamma _{T}$, so that
\begin{eqnarray}
\gamma _{F} &\equiv &[\phi _{F}(\tau _{\text{i}})-\phi _{F}(\tau _{\text{f}%
})]+\int F\cos (\theta _{F})d\phi _{F},  \nonumber \\
\gamma _{T} &\equiv &[\phi _{T}(\tau _{\text{i}})-\phi _{T}(\tau _{\text{f}%
})]+\int Fd\phi _{T},  \nonumber \\
\gamma  &=&\gamma _{F}+\gamma _{T}.
\end{eqnarray}%
From the definition, we see that $\gamma _{F}$ ($\gamma _{T}$) corresponds
to the rotation of the spin vector (tensor). As an example, we consider the
loop $\mathcal{L}$ in momentum space (by replacing the parameter $\tau $
with $\mathbf{k}$), and find that $\gamma _{F}=0$, $\gamma _{T}=\pi $. For a
small enough loop, the Berry phase $\gamma $ gives the local Berry flux
through the surface enclosed by the loop.

We want to emphasize that, to ensure a smooth wave function $|\Psi (\tau
)\rangle $, $F(\tau ),\phi _{F}(\tau ),\theta _{F}(\tau ),\phi _{T}(\tau )$
should also be a smooth function of $\tau $ except the points where $\langle
\mathbf{F}\rangle $ crosses the $z$-axis on or inside the Bloch sphere,
where $\phi _{F}(\tau ),\theta _{F}(\tau ),\phi _{T}(\tau )$ may have jumps.
We can simply remove these points in the integral that do not affect the
final results.

\subsection{D. Majorana star representation of Berry flux}
An arbitrary spin-1 quantum
state  can be
written as
$|\Psi \rangle =\sum_s f_s |s\rangle$,
and we will use $s=-1,0,1$ to
represent the spin state $\downarrow, 0,\uparrow$ for convenience.
We can rewrite the spin-1 basis using
the two-mode boson basis
with $|s\rangle=\frac{(c^\dag)^{1+s}(d^\dag)^{1-s}}{(1+s)!(1-s)!}|\emptyset\rangle$,
then the state can be factorized as $|\Psi \rangle=\frac{1}{\mathcal{N}}\prod^2_{j=1}\alpha_j^\dag|\emptyset\rangle$,
with $\mathcal{N}$ the normalization coefficient and $\alpha_j^\dag=\cos(\theta_j/2)c^\dag+\sin(\theta_j/2)e^{i\varphi_j}d^\dag$.
If we denote $c^\dag|\emptyset\rangle$ and $d^\dag|\emptyset\rangle$
as the spin-1/2 basis, then the above factorization will give out
2 pairs of parameters $(\theta_j,\varphi_j)$ which corresponds
to 2 Majorana stars $\mathbf{m}_j=(\sin\theta_j\cos\varphi_j,\sin\theta_j\cos\varphi_j,\cos\theta_j)$ on the Bloch sphere.
The parameters are determined by~\cite{PhysRevLett.113.240403}
$q_j=\tan(\theta_j/2)e^{i\varphi_j}$ with $q_1$ and $q_2$
the roots of the equation
$\sum_{j=0}^2 \frac{(-1)^jf_{1-j}}{\sqrt{j!(2-j)!}}q^{2-j}=0$.
The Berry phase accumulated along a loop
can be formulated as~\cite{PhysRevLett.113.240403,PhysRevA.98.013627}
\begin{eqnarray}
\gamma &=&i\int d\tau \langle \Psi (\tau )|\partial _{\tau }|\Psi (\tau
)\rangle +\gamma _{\text{f,i}} \nonumber \\
&=& -\frac{1}{2}\oint\frac{\mathbf{m}_1\times\mathbf{m}_2\cdot (d\mathbf{m}_1-d\mathbf{m}_2)}{3+\mathbf{m}_1\cdot\mathbf{m}_2} \nonumber \\
&&-\sum_{j=1}^2 \frac{1}{2}\oint (1-\cos\theta_j)d\varphi_j+\gamma _{\text{f,i}}.
\label{eq:Maj_S}
\end{eqnarray}%
The first term arises from the
correlations between the two Majorana stars and
the second term denotes the solid angles
traced out by them.
For the loop $\mathcal{L}$ in Fig.~\ref{fig:Berry}e,
$\gamma _{\text{f,i}}$ and the correlation term are both zero,
and the Berry flux is determined solely by the solid angle traced out by the two
Majorana stars.

\end{document}